\documentclass[%
 aps,
 superscriptaddress,
 nofootinbib,
 amsmath,amssymb,
 aps,
 prb,
 twocolumn,
]{revtex4-2}

\usepackage{dcolumn}
\usepackage{graphicx, amssymb, verbatim, amsmath, siunitx, color, float, soul, makecell, xcolor}
\usepackage[normalem]{ulem}
\usepackage[justification=raggedright,singlelinecheck=false]{caption}
\usepackage{natbib,url, enumerate}
\usepackage[colorlinks, allcolors= blue, bookmarksopen,bookmarksnumbered]{hyperref}

\definecolor{colsq}{rgb}{0,0.4470,0.7410}
\definecolor{coltr}{rgb}{0.8500,0.3250,0.0980}
\definecolor{coldm}{rgb}{0,0.4470,0.7410}
\definecolor{colbtr}{rgb}{0.9290,0.6940,0.1250}

\usepackage{amsfonts,bm}
\usepackage{placeins}
\usepackage{subfigure}
\usepackage{caption}
\usepackage{tabularx,ragged2e,booktabs,caption}
\usepackage{grffile}
\makeatletter
\newcommand*{\rom}[1]{\expandafter\@slowromancap\romannumeral #1@}
\makeatother

\def\beq{\begin{equation}}
\def\eeq{\end{equation}}
\def\ba{\begin{eqnarray}}
\def\ea{\end{eqnarray}}

\begin{document}

\preprint{APS/123-QED}

\title{Relaxation of the Random Site Coulomb glass Model in Two Dimensions}
\author{Preeti Bhandari}
 \affiliation{Faculty of Physical Sciences, South Asian University, New Delhi 110068, India.}
\affiliation{Faculty of Interdisciplinary Studies, South Asian University, New Delhi - 110068, India}
\author{Vikas Malik}
\email{vikasm76@gmail.com}
 \affiliation{Department of Physics and Material Science, Jaypee Institute of Information Technology, Uttar Pradesh 201309, India.}

\date{\today}

\begin{abstract}

 This study investigates the influence of material density, disorder in onsite energies, and localization length on relaxation dynamics within a two-dimensional random site Coulomb glass model at half-filling. To explore relaxation laws, we calculate the eigenvalue distribution of the linear dynamical matrix using mean-field approximations. Our findings indicate that the system initially undergoes rapid relaxation through energy-lowering transitions. The depletion of the single-particle density of states (DOS) near the Fermi level leads to slow relaxation, with fluctuations diminishing according to a power law. Subsequently, the system adheres to an exponential decay law after a specific period, defined as the relaxation time, which is inversely related to the minimum eigenvalue of the dynamical matrix. As the density of the system decreases, the relaxation rate slows down, resulting in an increase in the relaxation time. For a constant density and localization length, an increase in the disorder of onsite energies results in a longer relaxation time. A significant portion of the eigenvalue spectrum remains unaffected, suggesting that a reduction in localization length concurrent with increased disorder may play an equally vital role in the slow dynamics observed.
\end{abstract}


\maketitle

\section{Introduction}
 The Coulomb glass (CG) model \cite{shklovskii2013electronic,pollak2013electron} serves as a theoretical framework to describe systems of localized electrons that interact via an unscreened Coulomb potential. The sites can be positioned randomly in the system or on a lattice. Each site may also have a random energy associated with it. The Hamiltonian that governs the Coulomb glass system \cite{davies1982electron} is mathematically formulated as:
\begin{equation}
\label{Hamiltonian}
\mathcal{H} \{n_{i}\} = \sum_{i=1}^{N} \phi_{i} n_{i} +  \frac{1}{2} \sum_{i \neq j} \frac{e^{2}}{\kappa |\vec{r_{i}} - \vec{r_{j}}|} (n_{i} - 1/2) (n_{j} - 1/2).
\end{equation}  
Here, $\phi_{i}$ represents the random site energy at site $i$, and the occupation number $n_{i} \in \{0,1\}$. The electrons at site $i$ and $j$ separated by distance $r_{ij}$ interact through an unscreened Coulomb interaction $e^{2}/(\kappa \ r_{ij})$ where $\kappa$ is the dielectric constant. In the Coulomb glass lattice model, the sites are located in a regular lattice, and the disorder is incorporated through the $\phi_{i}$'s. In the random site CG model, the sites are randomly placed, leading to the interaction energy becoming a random variable, and $\phi_{i}$ may be non-zero. At absolute zero temperature, when the disorder and the interaction strength are of comparable magnitudes, a soft gap emerges in the single-particle density of states \cite{efros1975coulomb,shklovskii2013electronic,baranovskii1979coulomb,davies1984properties,mobius1992coulomb,glatz2008coulomb,bhandari2017critical,bhandari2017effect,goethe2009phase,surer2009density,mobius2010comment}. This gap is progressively filled as the temperature of the system increases \cite{sarvestani1995coulomb}.

The interplay of disorder and Coulomb interaction leads to glassy behavior in the CG system \cite{davies1982electron,grunewald1982mean,pollak1982coulomb,pollak1984non,grannan1994grannan,mueller2004glass,pastor1999melting,pankov2005nonlinear,mueller2007mean,bray1982spin,amir2009slow,amir2008mean,burin2006many}. This glassy behavior has been observed in numerical simulations \cite{bhandari2021relaxation,kolton2005heterogeneous,kirkengen2009slow} for both lattice and random site CG models. \cite{davies1982electron,grunewald1982mean,pollak1982coulomb,pollak1984non,grannan1994grannan,mueller2004glass,pastor1999melting,pankov2005nonlinear,mueller2007mean,bray1982spin,amir2009slow,amir2008mean,burin2006many}. In contrast, most experiments are conducted on random site models \cite{vaknin2000aging,vaknin2000heuristic,vaknin2002nonequilibrium,orlyanchik2004stress,kozub2008memory,ben1993nonequilibrium,martinez1997anomalous,martinez1998coulomb}, where the system is driven out of equilibrium using various perturbation techniques. The ensuing glassy dynamics manifests as a very slow relaxation back towards equilibrium: an initial rapid relaxation regime is followed by a much slower relaxation that is typically described by power-law or logarithmic time dependence. The origin of this slow relaxation and, more generally, of the glassy behavior remains a subject of active debate. The Coulomb gap is widely believed to play a central role in accounting for the experimental observations \cite{clare1999time,malik2004formation,lebanon2005memory}. Since the Coulomb gap is due to the long range nature of Coulomb interaction, the unscreened nature of Coulomb interaction is vital to the glassy dynamics scenario. However, both experiments \cite{ovadyahu2019screening} and simulations \cite{bhandari2021relaxation} that incorporate screened Coulomb interactions still exhibit slow relaxation, raising the question of the necessity and dominant role of unscreened Coulomb interactions. An alternative viewpoint attributes the primary origin of slow relaxation to a strong disorder \cite{ovadyahu2017disorder}.

To effectively explore the dynamics of these systems, one can utilize the generalized master equation \cite{Puri2009, bhandari2021relaxation}. The rate of change of probability $P (\lbrace n_{\mu}\rbrace, t)$ to find the system in state $\mu$ at time $t$ is,
\ba
\label{eq1}
\dfrac{d}{dt} P (\lbrace n_{\mu}\rbrace, t) = - \sum_{\mu \neq \nu}  \hspace*{2mm} W_{\mu \rightarrow \nu} \hspace*{2mm} P(\lbrace n_{\mu}\rbrace,t) \nonumber \\
+ \sum_{\nu \neq \mu}  \hspace*{2mm} W_{\nu \rightarrow \mu} \hspace*{2mm} P(\lbrace n_{\nu}\rbrace,t)
\ea
where $W_{\mu \rightarrow \nu}$ represents the transition rates from state $\mu$ to $\nu$. Transitions between states can occur via single-electron or multiple-electron transfer hops. The total number of particles (electrons) is conserved in CG systems, and thus only particle (electron) conserving transitions are considered. There are two ways to explain the glassy dynamics. In the first picture, an out of equilibrium system relaxes to a metastable state and is trapped due to the high energy barriers between metastable states. Consequently, this results in an initial rapid relaxation phase (relaxation to the metastable state minimum) followed by a prolonged slow decay (transition between metastable states). In contrast, the second picture only considers a single valley (metastable state). It shows that the system's relaxation towards a local minimum is slow when it is slightly away from the local equilibrium. This paper uses the second picture and assumes that only single-particle transitions are important. In the context of this approximation, we employ mean-field methods to describe the system. Within this formalism, the average occupation at a site $i$  at temperature $T$ is given by the Fermi-Dirac distribution,

\begin{equation}
			\label{mag}
			f_{i} = f_{FD}(\varepsilon_{i})=(1/(exp[\beta \varepsilon_{i}] + 1)  , 
		\end{equation}
where  $\varepsilon_i$ is the Hartree energy at site $i$ and $\beta=1/T$.  Defining $ J_{ik}= e^{2}/(\kappa \ r_{ij})$, $\varepsilon_{i}$ is defined as
		\begin{equation}
			\label{HE}
			\varepsilon_{i} = \phi_{i} + \sum_{k \neq i} J_{ik} (f_{k}-1/2) \ .
		\end{equation}
Solving Eq.(\ref{mag}) and Eq.(\ref{HE}) self-consistently gives the average occupation numbers for the system at a particular disorder configuration at a temperature $T$.

The average occupation at a site $i$  at a time $t$ in the mean field approximation, close to local equilibrium is
\begin{equation}
\label{eq15}
N_{i}(t) = f_{i} + \delta N_{i}
\end{equation}
 The time evolution of the fluctuation $\delta N_{i}(t)$  is controlled by the matrix equation
\begin{equation}
\dfrac{d}{dt} \delta N_{i}= \sum_{l} A_{i l} \hspace*{2mm} \delta N_{l}
\end{equation} 
where we define
\begin{subequations}
	\label{A-mat}
	\begin{equation}
	\Gamma_{ik} = \frac{1}{2 \tau_1} \gamma(r_{i k}) \hspace*{2mm} f_{i} (1-f_{k}) \hspace*{2mm} f_{FD}(\varepsilon_{k} - \varepsilon_{i})
	\end{equation}  
	\begin{equation}
	\Gamma_{ k i} = \frac{1}{2 \tau_1} \gamma(r_{k i}) \hspace*{2mm} f_{k} (1-f_{i}) \hspace*{2mm} f_{FD}(\varepsilon_{i} - \varepsilon_{k})
	\end{equation}
    \begin{equation}
	A_{i i}  = -\sum_{k \neq i}\, \frac{\Gamma_{i k}}{f_{i}(1-f_{i})} 
	\end{equation}
	\begin{equation}
	A_{i l} = \frac{\Gamma_{l i}}{f_{l}(1-f_{l})} + \frac{1}{T} \sum_{k(\neq l\neq i)}\, \Gamma_{i k} \hspace*{2mm} (J_{k l}-J_{i l})
	\end{equation}
\end{subequations}  
where $\gamma(r)=exp(-r/a)$ for the system that has localization length $a$. The equilibrium transition rates are represented by $\Gamma_{ik}$.  The `` A-matrix" eigenvalue distribution describes the system's dynamics, which is pushed marginally away from its local equilibrium state. 
In recent experiments \cite{ovadyahu2018transition} on  Anderson- insulating amorphous indium oxide $In_x O$ films, a crossover was observed from logarithmic decay $\delta G(t) \propto log(t)$ of excess conductance to the exponential decay law $\delta G(t) \propto exp(-t/\tau)$. Using samples with different degrees of disorder, $\tau$ is shown to vanish at the critical metal-insulator transition point. Thus, the system relaxes faster as the disorder in the system decreases. The film with high sheet resistance (high degree of disorder) was annealed to lower the sheet resistance (lower degree of disorder). The study of changes in the structural properties of the material during the annealing process revealed that the change in resistance is mainly due to an increase in the density of the material \cite{givan}. The investigated samples had the same carrier concentration and thus it was proposed that the decrease in their relaxation times as the sample resistance decreased was due to the decrease in the quenched disorder. 

Indium oxide films are amorphous and retain their amorphous character up to the metal-insulator transition point. In this paper, we focus on the random site model to model the amorphous nature of indium oxide films.  Experiments reveal that density of sites and disorder in the system are the important parameters affecting the relaxation of the system. First, we keep random onsite energies as zero and study the effect of changing the density of the materiel on the relaxation dynamics of the system. Thus, the disorder is only due to the random positioning of sites. Next, keeping the materials density constant, we consider the effect of increasing the disorder in the onsite energies on the relaxation dynamics of the system. Finally, we look at the effect of localization length on the relaxation dynamics and discuss its importance.

Our paper is organized as follows. In Sec. \ref{simulations} , we describe the details of the simulation. In Sec. \ref{resultSec}, we discuss our results.  Finally, Sec. \ref{conclude} concludes the paper with a summary of our results.

\section{Simulation Details}
 \label{simulations}
In this work, we consider a two-dimensional system of size $L \times L$ with $L=40$. The number of sites (N) are randomly placed in a square box of size L. The minimum distance between sites is kept at 0.05.  Defining $N=nL^2$, we work at $n=0.5, 1.0, 2.0$ in this study. The number of electrons is half the total number of sites in the system. The localization length is  $a=0.25$ and  $T=0.05$ except when mentioned otherwise. We use periodic boundary conditions. All quantities are averaged over 100 disorder configurations.  The random onsite energy are chosen from the probability distribution,
\begin{equation}
 \label{field_prob}
  P(\phi)=\begin{cases}
    \frac{1}{W}, & \text{if $-W/2\leq \phi \leq W/2$}.\\
    0, & \text{otherwise}.
  \end{cases}
\end{equation} 
Here, $W$ denotes the strength of onsite disorder.

 For each disorder configuration, we calculate $f_i$ and  $\varepsilon_i$  using Eq. (\ref{mag}) and Eq.(\ref{HE}).  The corresponding transition rates were determined between all pairs of sites. The matrix elements of the dynamical matrix $A$ were then calculated and the matrix $A$ was diagonalized to find its eigenvectors and eigenvalues $(\lambda)$.

\section{Results}
\label{resultSec}
\begin{figure}
\centering
\includegraphics[scale=0.3]{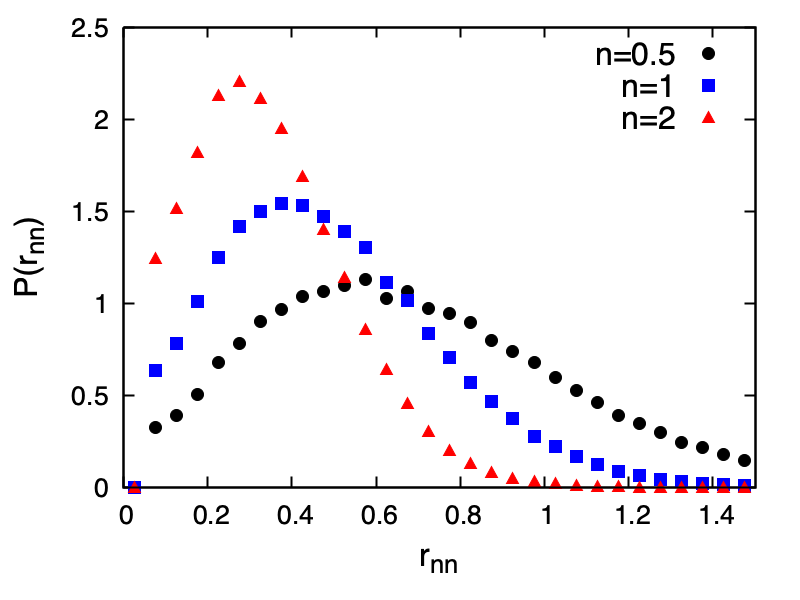}
\caption{Probability density of $r_{nn}$, where $r_{nn}$ is defined in the text.}
\label{rnn}
\end{figure}

 First, we study the effect of density variation on the relaxation of fluctuations for the random site CG model with zero onsite energies. Since the sites are randomly placed, the interaction energy $J_{ik}$ is a random quantity and is the only source of disorder in the system. To calculate the disorder, we first identify the nearest neighbor for each site and calculate the distance ($r_{nn}$) between them. The probability distribution for $r_{nn}$ at different densities is shown in Fig.(\ref{rnn}). As the density of the sites increases, the average distance between the sites decreases. The width of the probability distribution increases as $n$ decreases.

\begin{figure}
\centering

\includegraphics[scale=0.3]{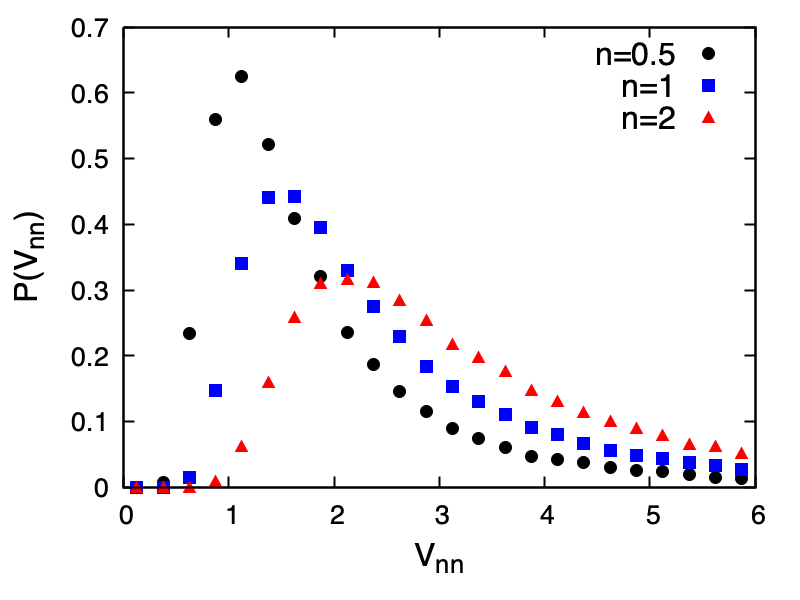}
\caption{Probability density of $V_{nn}=1/r_{nn}$.} 
\label{Vnn}
\end{figure}

 Defining $V_{nn}=1/r_{nn}$, the probability distribution of $V_{nn}$ is shown in Fig.(\ref{Vnn}). As the value of $n$ increases, the average value of $V_{nn}$ increases, implying that the strength of the interaction increases. We also observe that the width of the $P(V_{nn})$ distribution increases as $n$ increases, which implies that the strength of disorder in the system increases.
 
\begin{figure}
\centering
\includegraphics[scale=0.3]{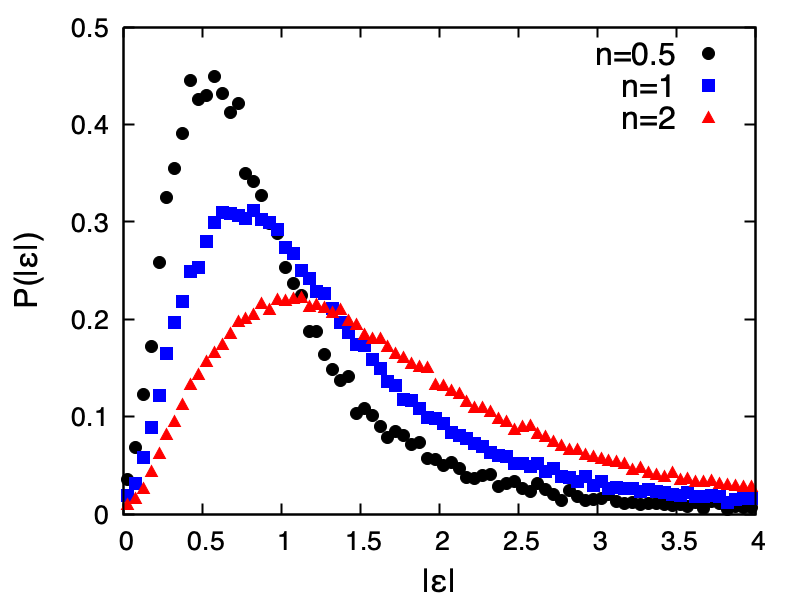}
\caption{Density of states for single particle energy  vs $|\varepsilon|$ for different electron density.}
\label{dos}
\end{figure}

The density of states $P(\varepsilon)$ for single particle Hartree energies defined in Eq.(\ref{HE}) is shown in Fig.(\ref{dos}) for different densities. For all values of $n$ considered, there is a soft gap in the density of states. The width of the gap and  $P(\varepsilon)$ increases as the density ($n$) increases. The width of the gap and $P(\varepsilon)$ are measures of the strength of the interaction and disorder, respectively. The results are consistent with the probability distribution for $V_{nn}$.

\begin{figure}
\centering
\includegraphics[scale=0.3]{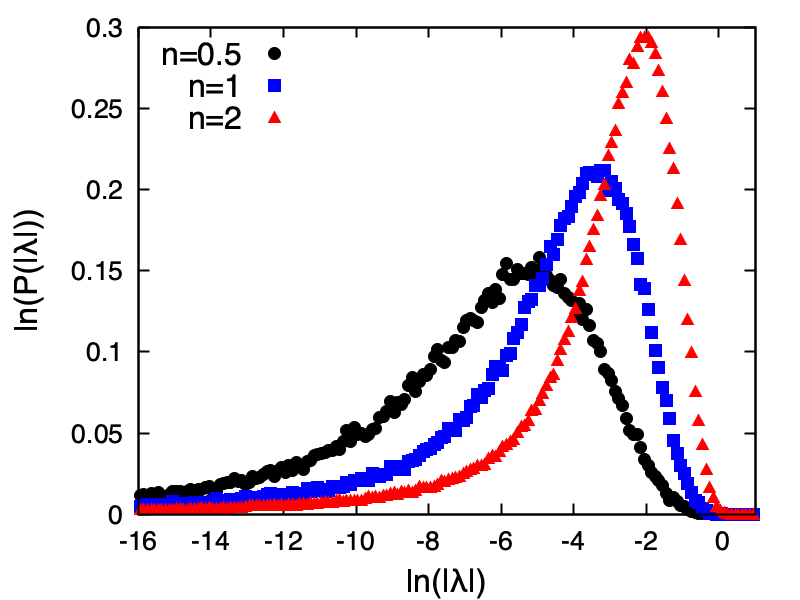}
\caption{Probability density of $\ln({|\lambda|})$ for $a=0.25$ and $T=0.05$ at different densities $n=0.5,1.0,2.0$.}
\label{Plambda}
\end{figure}

\begin{figure}
\centering
\includegraphics[scale=0.3]{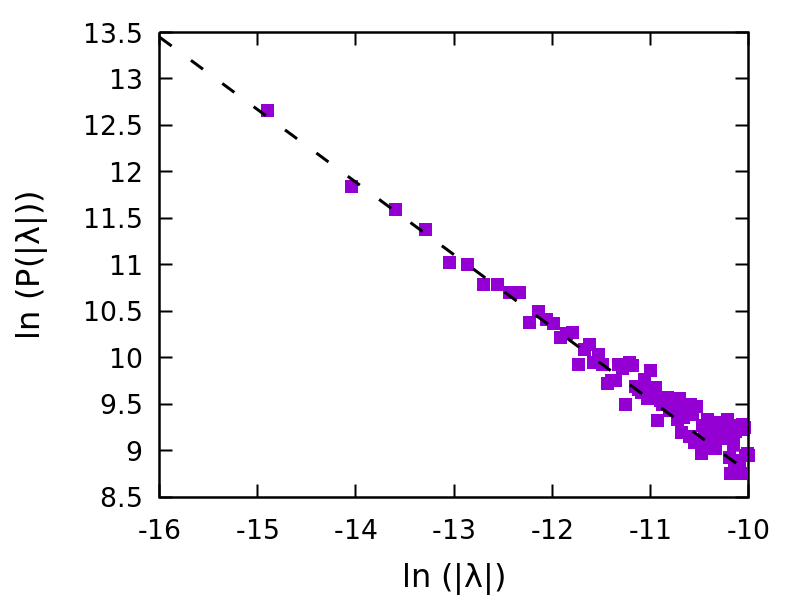}
\caption{ $\ln(P(|\lambda|))$  vs $\ln|\lambda|$ for $a=0.25$, $T=0.05$ at $n=1/2$ for the range $e^{-15}$ - $e^{-12}$. The graph is fitted to a straight line f(x)=ax+b, using a=-0.78 and b=0.96.}
\label{fitPlot}
\end{figure}

In experimental studies of relaxation, the system is first quenched from a high temperature to the desired low temperature and allowed to equilibrate for a long time. After equilibrium is reached, it is briefly driven out of equilibrium—typically for a short interval known as the waiting time—by applying an external perturbation. Once this perturbation is turned off, the system exhibits an initial fast relaxation, followed by a much slower decay. In the regime where this slow relaxation occurs, the response often follows a power-law or logarithmic behavior. Beyond a characteristic time scale $\tau$, relaxation typically transitions to exponential decay $\delta N(t) = exp(-t/\tau)$, as reported in many experiments. In our calculation, the relaxation of a disordered system slightly away from equilibrium depends on the eigenvalue spectrum of the dynamical matrix A defined in Eq. (\ref{A-mat}).  For $P(\lambda) \sim 1/\lambda$, the system relaxes via the logarithmic decay law ($\delta N(t) \sim log(t))$. The system shows a power lay decay, $\delta N(t) \sim 1/t^{1-\alpha}$ for $P(\lambda) \sim \lambda^{-\alpha}$. After a time $\tau =1/\lambda_{min}$, the relaxation of the system shows an exponential decay. The variation of $P(ln(|\lambda|))$ for $a=0.25$ and $T=0.05$ is shown in Fig. \ref{Plambda} for different densities.  The minimum eigenvalue $(\lambda_{min})$ of the matrix $A$ increases and therefore the relaxation time ($\tau = 1/\lambda_{min}$) decreases as the density increases. We calculate the probability distribution $P(\lambda)$ for the eigenvalues to find the functional form of the relaxation law at late times. In Fig. \ref{fitPlot},  the fitting of $ln(P(|\lambda|))$ versus $ln(|\lambda|)$ for small $\lambda$ at $n=1/2$ is shown. The fit is linear and thus $P(\lambda) \sim \lambda^{-\alpha}$, which implies that the fluctuations will decay according to the power law $\delta N(t) \sim 1/t^{1-\alpha}$.  A similar analysis for other densities shows that the exponent $\alpha$ decreases as the density increases.  We find $\alpha=0.78, 0.7185$ and $0.67$  for $n=1/2, 1$ and $2$, respectively. Therefore, the decay slows as the density decreases.

\begin{figure}
\centering
\includegraphics[scale=0.3]{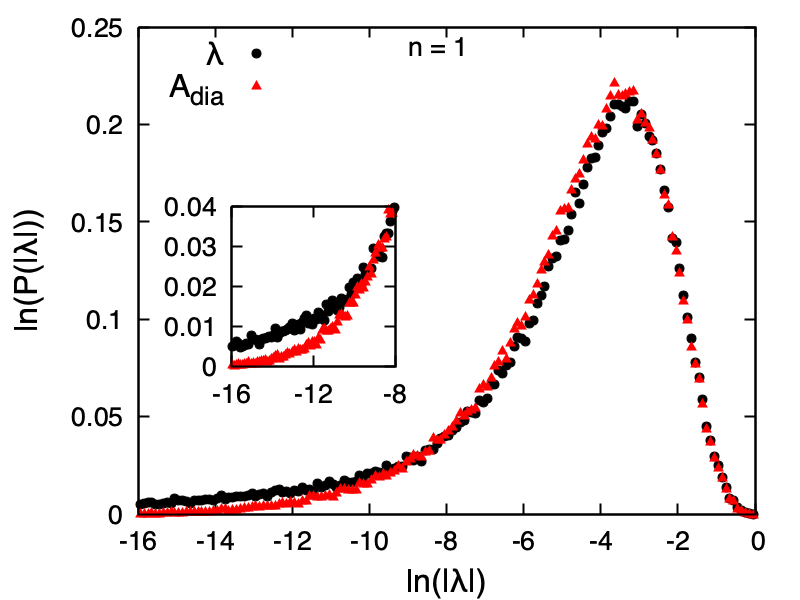}
\caption{Comparison of probability density of the eigenvalues of the A-matrix and its diagonal part $(A_{ii})$ for $a=0.25$ and $T=0.05$ at $n=1.0$. (inset) Data shown for $|\lambda|$ near $|\lambda_{min}|$.}
\label{compPlot}
\end{figure}

\begin{figure}
\centering
\includegraphics[scale=0.3]{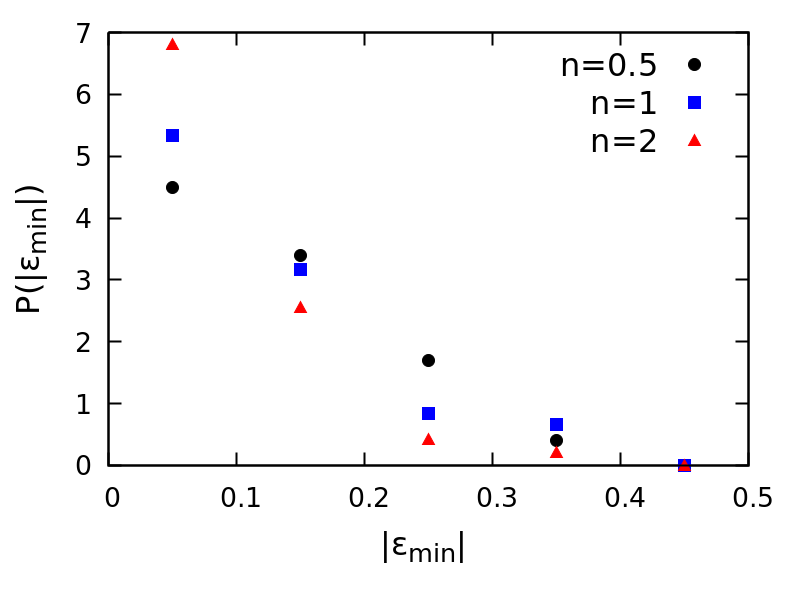}
\caption{Probability density of $\varepsilon_{min}$ for $a=0.25$ and $T=0.05$ at  different densities $n=0.5,1.0,2.0$.}
\label{ProbDense}
\end{figure}

To explain these results, we analyze the diagonal part of the $A$ matrix. The comparison of the distribution for $A_{ii}$ and $\lambda$ is made for $n=1.0$ in Fig. \ref{compPlot}. The distribution for $A_{ii}$ and $\lambda$  is almost identical except for the eigenvalues near $\lambda_{min}$.  For half filling and zero onsite disorder, the single particle density of states is symmetric around the Fermi level at $\varepsilon =0$. For $|\varepsilon_i| > T$ and $|\varepsilon_k|> T$, the transition rate $\Gamma_{ik}$ can be expressed as
\begin{equation}
	\label{Tik}
	\Gamma_{ik} = \frac{1}{2 \tau_1} \gamma(r_{i k}) \hspace*{2mm} exp[ -\frac{1}{2 T} (|\varepsilon_i|+|\varepsilon_k|+|\varepsilon_i - \varepsilon_k|) ]
\end{equation} 
Similarly, the factor $f_i(1-f_i) \sim exp(-\beta |\varepsilon_i|)$ for $\beta|\varepsilon_i| > 1$. Next, we analyze the contribution of the term $\Gamma_{ik}/f_i(1-f_i)$ to $A_{ii}$ for a site with Hartree energy $\varepsilon_i < 0$ such that $\beta|\varepsilon_i| > 1$. The energy dependent part of $\Gamma_{ik}/f_i(1-f_i)$ is denoted by $W_{ik}$ and can be approximately as 
\begin{equation}
 \label{Aii}
   W_{ik}=\begin{cases}
    1 & |\varepsilon_k| < |\varepsilon_i|, \varepsilon_k <0,\\
    e^{-\beta(|\varepsilon_k|-|\varepsilon_i|)} & |\varepsilon_k| > |\varepsilon_i|, \varepsilon_k <0,\\
     e^{-\beta|\varepsilon_k|} & \varepsilon_k >0
  \end{cases} 
 \end{equation}
So, a fluctuation at a site $i$ with $\beta|\varepsilon_i| > 1$, will decay the fastest by lowering its energy and jumping towards the Fermi level in terms of energy. The most probable jump will be for which the $W_{ik}\gamma_{ik}$ factor is maximized. For a site just below the Fermi level, an energy lowering jump has a small probability because of a Coulomb gap. Thus,  the decay of fluctuation will require energy, which will decrease the value of $A_{ii}$. Similar analysis applies to sites above the Fermi level. Thus, the value of $A_{ii}$ should be the smallest for sites near the Fermi level. To substantiate this, we identified the minimum value of the diagonal component of the dynamical matrix $A$ for each disorder configuration. For the corresponding site, we computed its Hartree energy, denoted as $\varepsilon_{min}$, using Eq. (\ref{HE}).  Fig. \ref{ProbDense} illustrates the probability distribution of $\varepsilon_{min}$ and shows that the smallest values for $A_{ii}$ are for sites close to the Fermi level.  

We can now understand the various features of the distribution for $ln(|\lambda|)$ shown in Fig. \ref{Plambda} and Fig. \ref{compPlot}. As the density increases, the maximum possible value of $\lambda$ or $A_{ii}$ shifts towards the right. For the maximum value of $A_{ii}$, the fluctuation at the site will make an energy lowering transition so that $W_{ik}=1$ and $A_{ii}$ are approximately equal to $\gamma_{ik}$. As density increases, $r_{av} \sim n^{-1/2}$ decreases and $\gamma_{ik}$ increases, and thus the maximum value of $\lambda$ or $A_{ii}$ increases. Since $\gamma_{ik}$ plays an important role in determining the eigenvalues, the distribution for $r_{nn}$ shown in Fig. \ref{rnn} affects the spectrum of eigenvalues. The width of the distribution for $r_{nn}$ is the minimum for $n=2$ and increases as the density decreases. The same trend is followed for the width of the distribution of $ln(|\lambda|)$. As the density increases, the Coulomb gap's width and the system's disorder increase. This does not lead to a slowing of decay with an increase in density, but, on the contrary, the fluctuation dissipates by jumping to a slightly larger distance to minimize the factor $r/a+\Delta E/T$. 

\begin{figure}
\centering
\includegraphics[scale=0.3]{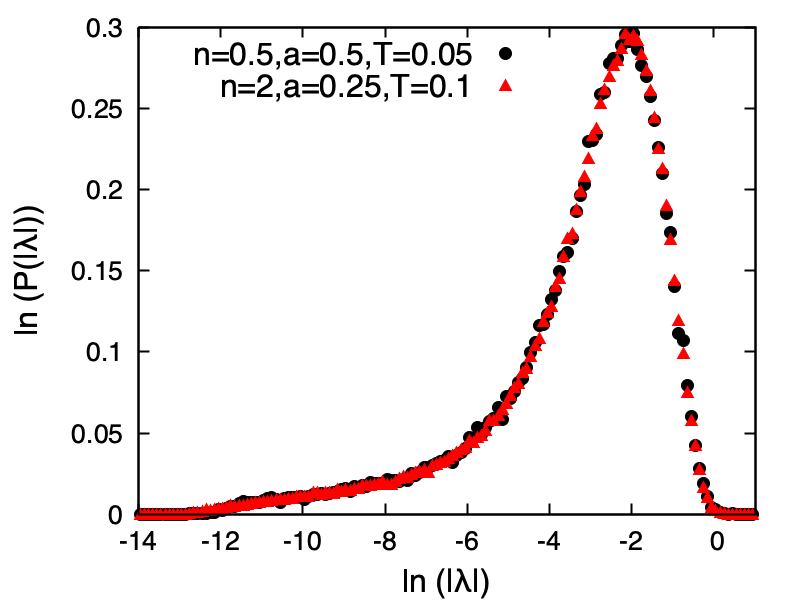}
\caption{Comparison of probability density of $ln({|\lambda}|)$ for $n=0.5, a=0.5, T=0.05$ with the $n=2$, $a=025$, $T=0.1$.}
\label{compareProb}
\end{figure}

\begin{figure}
\centering
\includegraphics[scale=0.3]{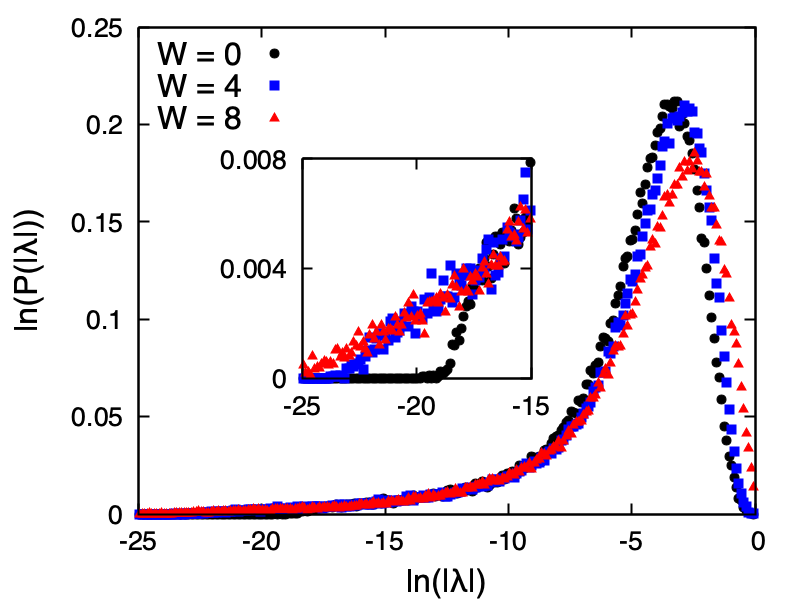}
\caption{Comparison of probability density of $\ln({|\lambda}|)$ for different disorders $W=0.0,4.0,8.0$ at $n=1$. (inset) Data shown for $|\lambda|$ near $|\lambda_{min}|$.}
\label{CompLambda}
\end{figure}

Our analysis shows that for $W=0$, the relaxation should be a function of dimensionless parameters, $r_{av}/a$ and $V_{nn}/T$.  We verify this scaling property by comparing the $n=0.5, a=0.5, T=0.05$ case with the $n=2, a=0.25, T=0.1$ case in Fig. \ref{compareProb}. In terms of density $(n)$, $r_{av}/a=c_1/n^{1/2} a$ and $V_{nn}/T=c_2n^{1/2}/T$, where $c_1$ and $c_2$ are constants. For the two cases compared in Fig. \ref{compareProb}, the values of $r_{av}/a$ and $V_{nn}/T$ are the same. The eigenvalue distribution is almost identical, indicating that the scaling is successful.  

\begin{figure}
\centering
\includegraphics[scale=0.3]{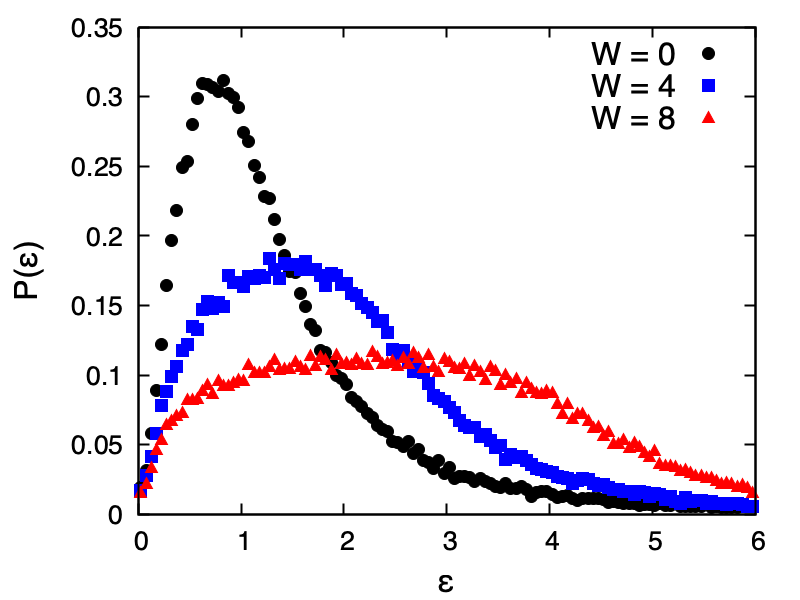}
\caption{Density of states for single particle energy  vs $|\epsilon|$ for different disorder strengths $W=0.0,4.0,8.0$ at $n=1$.} 
\label{dosW}
\end{figure}

\begin{figure}
\centering
\includegraphics[scale=0.3]{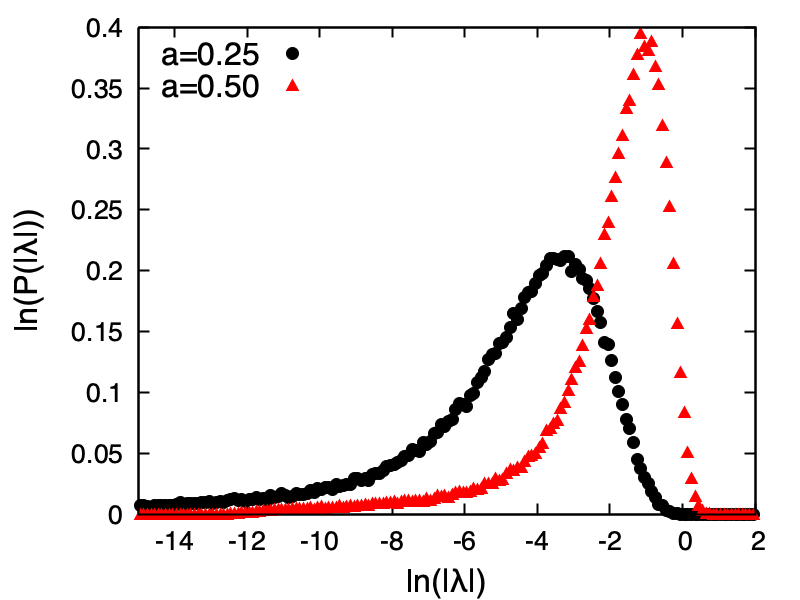}
\caption{Comparison of probability density of $\ln(|{\lambda})|$ for different localization lengths $a=0.25$ and $0.5$ at $n=1$, $W=0.0$.}
\label{localcomp}
\end{figure}

We now examine the effect of the disorder in onsite energies on the relaxation dynamics of the system at constant $n$. For relaxation effects, $P(\lambda)$ at small values of $\lambda$ is significant. Fig. \ref{CompLambda} shows that the probability distribution for small eigenvalues, $\lambda$ varies very little with disorder, except near $\lambda_{min}$. So, although the final decay to exponential decay may take longer as $\tau=1/ \lambda_{min}$ is larger for higher disorders, for most times, relaxation will be the same.  In experiments, the system is perturbed away from equilibrium for a time $t_w << \tau$ by an external source. Just after the perturbation is removed, the excess conductance  is higher for the system having larger disorder. Our calculation predicts that the relaxation at short times and hence the excess conductance is independent of the strength of disorder, which contradicts the experimental results \cite{ovadyahu1997disorder, ovadyahu2018transition}. After the perturbation has been removed, the relaxation rate is faster for a system with smaller disorder.  Our calculations again contradict the experimental results \cite{ovadyahu1997disorder, ovadyahu2018transition} and predict that the relaxation will be different only in the tail but nearly the same at other times.  The density of states for single particle energies for $n=1$ and different disorder strengths is shown in Fig. \ref{dosW}. One sees that as the disorder in onsite energies increases, the width of the Coulomb gap decreases, and the width of the distribution $P(\epsilon)$ increases. An increase in disorder leads to a significant decrease in conductivity in experiments and simulations. Mapping the Coulomb glass problem to a resistor network \cite{shklovskii2013electronic, pollak2013electron}, it has been shown that the conductivity of the system depends upon the backbone of the percolation network. On the other hand, slow relaxation is due to the sites in isolated clusters \cite{ovadyahu2017disorder}.  Our results support the notion that the sites in the isolated cluster contribute to the slow decay. We further propose that as the disorder in the system increases, the localization length of the system decreases. For constant values of $n=1$ and $W=0$, the probability distribution of $ln(|\lambda|)$ is shown in Fig. \ref{localcomp} for $a=0.25$ and $0.50$. The shift in the probability distribution towards the higher values of $\lambda$  implies that the decay for a higher value of $a$ is faster on all time scales. The value of $\tau$ also decreases as the localization length increases. Similar results were obtained for different densities and disorder strengths. This is expected since the transition to any site becomes faster as the localization length increases.

\section{Conclusions}
\label{conclude}
In summary, this study employs a random-site model to represent the amorphous character of the system and incorporates random onsite energies to account for the quenched disorder in amorphous indium oxide films. This theoretical framework is closely connected to recent experimental investigations of Anderson-insulating amorphous indium oxide ($\mathrm{In}_x\mathrm{O}$) films \cite{ovadyahu2018transition}, which have identified a crossover in the relaxation dynamics of the excess conductance from a logarithmic dependence, $\delta G(t) \propto \log(t)$, to an exponential decay, $\delta G(t) \propto \exp(-t/\tau)$. Thermal annealing of the samples accelerates this relaxation and leads to a reduction in the characteristic relaxation time $\tau$. Notably, annealing increases the mass density of the films and diminishes the degree of quenched disorder, while leaving the carrier concentration and the underlying amorphous structure essentially unchanged.

The influence of onsite-energy disorder and site density on relaxation dynamics can be summarized as follows. We begin by setting the onsite-energy disorder to zero and varying only the site density. As the density is lowered, the Coulomb gap becomes narrower, and the distribution of Hartree energies decreases accordingly. This leads to a reduction in both the effective interaction strength and the interaction-induced contribution to the disorder as the density decreases. Our results indicate that relaxation is slowest at the highest site density, where the dynamics is primarily governed by sites lying within the Coulomb gap.

The transition rate includes two factors: $\exp(-r/a)$, which represents the tunneling probability (with $r$ the intersite distance and $a$ the localization length), and $\exp(-\Delta E/T)$, which describes phonon-assisted hopping, where $\Delta E$ is the energy difference and $T$ the temperature. As the density increases, the average intersite distance decreases, and a fluctuation can relax by hopping to a slightly more distant site, thereby compensating for the increase in $\Delta E$ associated with the enhanced disorder strength and the stronger depletion of the density of states in the vicinity of the Fermi level. Given that the experimentally observed increase in density upon annealing is on the order of $10\%$, it is unlikely that this relatively modest change in density alone can account for the substantially accelerated relaxation. 

Next, we examine the role of quenched disorder in onsite energies $\phi_i$, in determining relaxation dynamics while keeping the density of the site fixed. We find that the relaxation time $\tau$ increases with increasing disorder strength; however, the relaxation curves coincide over most of the temporal evolution. Thus, a change in the strength of the disorder cannot alone explain the experiential results. We suggest that a reduction in quenched disorder enhances the localization length, thus increasing the tunneling probability and resulting in faster relaxation, in qualitative agreement with the experimental observations on annealed amorphous indium oxide films.

The conclusion is that since the direct effect of reduced site energy inhomogeneity is ruled out as the cause of the increased rate after annealing and the reduction in tunneling distances is sort of ruled out an increase in localization length remains as the most plausible mechanism.

 \section*{Acknowledgments}

 V.M. acknowledges funding from the Anusandhan National Research Foundation (SERB), India, under Research Grant No. CRG/2022/004029. We wish to thank the high-performance computing facility `Ramanujam Universe' at JIIT Noida for the computational facility. The authors thank  Z. Ovadyahu for the illuminating discussions.


\begin{thebibliography}{46}%
	\makeatletter
	\providecommand \@ifxundefined [1]{%
		\@ifx{#1\undefined}
	}%
	\providecommand \@ifnum [1]{%
		\ifnum #1\expandafter \@firstoftwo
		\else \expandafter \@secondoftwo
		\fi
	}%
	\providecommand \@ifx [1]{%
		\ifx #1\expandafter \@firstoftwo
		\else \expandafter \@secondoftwo
		\fi
	}%
	\providecommand \natexlab [1]{#1}%
	\providecommand \enquote  [1]{``#1''}%
	\providecommand \bibnamefont  [1]{#1}%
	\providecommand \bibfnamefont [1]{#1}%
	\providecommand \citenamefont [1]{#1}%
	\providecommand \href@noop [0]{\@secondoftwo}%
	\providecommand \href [0]{\begingroup \@sanitize@url \@href}%
	\providecommand \@href[1]{\@@startlink{#1}\@@href}%
	\providecommand \@@href[1]{\endgroup#1\@@endlink}%
	\providecommand \@sanitize@url [0]{\catcode `\\12\catcode `\$12\catcode
		`\&12\catcode `\#12\catcode `\^12\catcode `\_12\catcode `\%12\relax}%
	\providecommand \@@startlink[1]{}%
	\providecommand \@@endlink[0]{}%
	\providecommand \url  [0]{\begingroup\@sanitize@url \@url }%
	\providecommand \@url [1]{\endgroup\@href {#1}{\urlprefix }}%
	\providecommand \urlprefix  [0]{URL }%
	\providecommand \Eprint [0]{\href }%
	\providecommand \doibase [0]{https://doi.org/}%
	\providecommand \selectlanguage [0]{\@gobble}%
	\providecommand \bibinfo  [0]{\@secondoftwo}%
	\providecommand \bibfield  [0]{\@secondoftwo}%
	\providecommand \translation [1]{[#1]}%
	\providecommand \BibitemOpen [0]{}%
	\providecommand \bibitemStop [0]{}%
	\providecommand \bibitemNoStop [0]{.\EOS\space}%
	\providecommand \EOS [0]{\spacefactor3000\relax}%
	\providecommand \BibitemShut  [1]{\csname bibitem#1\endcsname}%
	\let\auto@bib@innerbib\@empty
	\bibitem [{\citenamefont {Shklovskii}\ and\ \citenamefont
		{Efros}(2013)}]{shklovskii2013electronic}%
	\BibitemOpen
	\bibfield  {author} {\bibinfo {author} {\bibfnamefont {B.~I.}\ \bibnamefont
			{Shklovskii}}\ and\ \bibinfo {author} {\bibfnamefont {A.~L.}\ \bibnamefont
			{Efros}},\ }\href@noop {} {\emph {\bibinfo {title} {Electronic properties of
				doped semiconductors}}},\ Vol.~\bibinfo {volume} {45}\ (\bibinfo  {publisher}
	{Springer Science \& Business Media},\ \bibinfo {year} {2013})\BibitemShut
	{NoStop}%
	\bibitem [{\citenamefont {Pollak}\ \emph {et~al.}(2013)\citenamefont {Pollak},
		\citenamefont {Ortu{\~n}o},\ and\ \citenamefont
		{Frydman}}]{pollak2013electron}%
	\BibitemOpen
	\bibfield  {author} {\bibinfo {author} {\bibfnamefont {M.}~\bibnamefont
			{Pollak}}, \bibinfo {author} {\bibfnamefont {M.}~\bibnamefont {Ortu{\~n}o}},\
		and\ \bibinfo {author} {\bibfnamefont {A.}~\bibnamefont {Frydman}},\
	}\href@noop {} {\emph {\bibinfo {title} {The electron glass}}}\ (\bibinfo
	{publisher} {Cambridge University Press},\ \bibinfo {year}
	{2013})\BibitemShut {NoStop}%
	\bibitem [{\citenamefont {Davies}\ \emph {et~al.}(1982)\citenamefont {Davies},
		\citenamefont {Lee},\ and\ \citenamefont {Rice}}]{davies1982electron}%
	\BibitemOpen
	\bibfield  {author} {\bibinfo {author} {\bibfnamefont {J.}~\bibnamefont
			{Davies}}, \bibinfo {author} {\bibfnamefont {P.}~\bibnamefont {Lee}},\ and\
		\bibinfo {author} {\bibfnamefont {T.}~\bibnamefont {Rice}},\ }\bibfield
	{title} {\bibinfo {title} {Electron glass},\ }\href@noop {} {\bibfield
		{journal} {\bibinfo  {journal} {Phys. Rev. Lett.}\ }\textbf {\bibinfo
			{volume} {49}},\ \bibinfo {pages} {758} (\bibinfo {year} {1982})}\BibitemShut
	{NoStop}%
	\bibitem [{\citenamefont {{\'E}fros}\ and\ \citenamefont
		{Shklovskii}(1975)}]{efros1975coulomb}%
	\BibitemOpen
	\bibfield  {author} {\bibinfo {author} {\bibfnamefont {A.~L.}\ \bibnamefont
			{{\'E}fros}}\ and\ \bibinfo {author} {\bibfnamefont {B.~I.}\ \bibnamefont
			{Shklovskii}},\ }\bibfield  {title} {\bibinfo {title} {Coulomb gap and low
			temperature conductivity of disordered systems},\ }\href@noop {} {\bibfield
		{journal} {\bibinfo  {journal} {Journal of Physics C: Solid State Physics}\
		}\textbf {\bibinfo {volume} {8}},\ \bibinfo {pages} {L49} (\bibinfo {year}
		{1975})}\BibitemShut {NoStop}%
	\bibitem [{\citenamefont {Baranovskii}\ \emph {et~al.}(1979)\citenamefont
		{Baranovskii}, \citenamefont {Efros}, \citenamefont {Gelmont},\ and\
		\citenamefont {Shklovskii}}]{baranovskii1979coulomb}%
	\BibitemOpen
	\bibfield  {author} {\bibinfo {author} {\bibfnamefont {S.}~\bibnamefont
			{Baranovskii}}, \bibinfo {author} {\bibfnamefont {A.}~\bibnamefont {Efros}},
		\bibinfo {author} {\bibfnamefont {B.}~\bibnamefont {Gelmont}},\ and\ \bibinfo
		{author} {\bibfnamefont {B.}~\bibnamefont {Shklovskii}},\ }\bibfield  {title}
	{\bibinfo {title} {Coulomb gap in disordered systems: computer simulation},\
	}\href@noop {} {\bibfield  {journal} {\bibinfo  {journal} {Journal of Physics
				C: Solid State Physics}\ }\textbf {\bibinfo {volume} {12}},\ \bibinfo {pages}
		{1023} (\bibinfo {year} {1979})}\BibitemShut {NoStop}%
	\bibitem [{\citenamefont {Davies}\ \emph {et~al.}(1984)\citenamefont {Davies},
		\citenamefont {Lee},\ and\ \citenamefont {Rice}}]{davies1984properties}%
	\BibitemOpen
	\bibfield  {author} {\bibinfo {author} {\bibfnamefont {J.}~\bibnamefont
			{Davies}}, \bibinfo {author} {\bibfnamefont {P.}~\bibnamefont {Lee}},\ and\
		\bibinfo {author} {\bibfnamefont {T.}~\bibnamefont {Rice}},\ }\bibfield
	{title} {\bibinfo {title} {Properties of the electron glass},\ }\href@noop {}
	{\bibfield  {journal} {\bibinfo  {journal} {Phys. Rev. B}\ }\textbf {\bibinfo
			{volume} {29}},\ \bibinfo {pages} {4260} (\bibinfo {year}
		{1984})}\BibitemShut {NoStop}%
	\bibitem [{\citenamefont {M{\"o}bius}\ \emph {et~al.}(1992)\citenamefont
		{M{\"o}bius}, \citenamefont {Richter},\ and\ \citenamefont
		{Drittler}}]{mobius1992coulomb}%
	\BibitemOpen
	\bibfield  {author} {\bibinfo {author} {\bibfnamefont {A.}~\bibnamefont
			{M{\"o}bius}}, \bibinfo {author} {\bibfnamefont {M.}~\bibnamefont
			{Richter}},\ and\ \bibinfo {author} {\bibfnamefont {B.}~\bibnamefont
			{Drittler}},\ }\bibfield  {title} {\bibinfo {title} {Coulomb gap in two-and
			three-dimensional systems: Simulation results for large samples},\
	}\href@noop {} {\bibfield  {journal} {\bibinfo  {journal} {Phys. Rev. B}\
		}\textbf {\bibinfo {volume} {45}},\ \bibinfo {pages} {11568} (\bibinfo {year}
		{1992})}\BibitemShut {NoStop}%
	\bibitem [{\citenamefont {Glatz}\ \emph {et~al.}(2008)\citenamefont {Glatz},
		\citenamefont {Vinokur}, \citenamefont {Bergli}, \citenamefont {Kirkengen},\
		and\ \citenamefont {Galperin}}]{glatz2008coulomb}%
	\BibitemOpen
	\bibfield  {author} {\bibinfo {author} {\bibfnamefont {A.}~\bibnamefont
			{Glatz}}, \bibinfo {author} {\bibfnamefont {V.~M.}\ \bibnamefont {Vinokur}},
		\bibinfo {author} {\bibfnamefont {J.}~\bibnamefont {Bergli}}, \bibinfo
		{author} {\bibfnamefont {M.}~\bibnamefont {Kirkengen}},\ and\ \bibinfo
		{author} {\bibfnamefont {Y.~M.}\ \bibnamefont {Galperin}},\ }\bibfield
	{title} {\bibinfo {title} {The coulomb gap and low energy statistics for
			coulomb glasses},\ }\href@noop {} {\bibfield  {journal} {\bibinfo  {journal}
			{Journal of Statistical Mechanics: Theory and Experiment}\ }\textbf {\bibinfo
			{volume} {2008}},\ \bibinfo {pages} {P06006} (\bibinfo {year}
		{2008})}\BibitemShut {NoStop}%
	\bibitem [{\citenamefont {Bhandari}\ \emph {et~al.}(2017)\citenamefont
		{Bhandari}, \citenamefont {Malik},\ and\ \citenamefont
		{Ahmad}}]{bhandari2017critical}%
	\BibitemOpen
	\bibfield  {author} {\bibinfo {author} {\bibfnamefont {P.}~\bibnamefont
			{Bhandari}}, \bibinfo {author} {\bibfnamefont {V.}~\bibnamefont {Malik}},\
		and\ \bibinfo {author} {\bibfnamefont {S.~R.}\ \bibnamefont {Ahmad}},\
	}\bibfield  {title} {\bibinfo {title} {Critical behavior of the
			two-dimensional coulomb glass at zero temperature},\ }\href@noop {}
	{\bibfield  {journal} {\bibinfo  {journal} {Phys. Rev. B}\ }\textbf {\bibinfo
			{volume} {95}},\ \bibinfo {pages} {184203} (\bibinfo {year}
		{2017})}\BibitemShut {NoStop}%
	\bibitem [{\citenamefont {Bhandari}\ and\ \citenamefont
		{Malik}(2017)}]{bhandari2017effect}%
	\BibitemOpen
	\bibfield  {author} {\bibinfo {author} {\bibfnamefont {P.}~\bibnamefont
			{Bhandari}}\ and\ \bibinfo {author} {\bibfnamefont {V.}~\bibnamefont
			{Malik}},\ }\bibfield  {title} {\bibinfo {title} {Effect of increasing
			disorder on domains of the 2d coulomb glass},\ }\href@noop {} {\bibfield
		{journal} {\bibinfo  {journal} {Journal of Physics: Condensed Matter}\
		}\textbf {\bibinfo {volume} {29}},\ \bibinfo {pages} {485402} (\bibinfo
		{year} {2017})}\BibitemShut {NoStop}%
	\bibitem [{\citenamefont {Goethe}\ and\ \citenamefont
		{Palassini}(2009)}]{goethe2009phase}%
	\BibitemOpen
	\bibfield  {author} {\bibinfo {author} {\bibfnamefont {M.}~\bibnamefont
			{Goethe}}\ and\ \bibinfo {author} {\bibfnamefont {M.}~\bibnamefont
			{Palassini}},\ }\bibfield  {title} {\bibinfo {title} {Phase diagram,
			correlation gap, and critical properties of the coulomb glass},\ }\href@noop
	{} {\bibfield  {journal} {\bibinfo  {journal} {Phys. Rev. Lett.}\ }\textbf
		{\bibinfo {volume} {103}},\ \bibinfo {pages} {045702} (\bibinfo {year}
		{2009})}\BibitemShut {NoStop}%
	\bibitem [{\citenamefont {Surer}\ \emph {et~al.}(2009)\citenamefont {Surer},
		\citenamefont {Katzgraber}, \citenamefont {Zimanyi}, \citenamefont
		{Allgood},\ and\ \citenamefont {Blatter}}]{surer2009density}%
	\BibitemOpen
	\bibfield  {author} {\bibinfo {author} {\bibfnamefont {B.}~\bibnamefont
			{Surer}}, \bibinfo {author} {\bibfnamefont {H.~G.}\ \bibnamefont
			{Katzgraber}}, \bibinfo {author} {\bibfnamefont {G.~T.}\ \bibnamefont
			{Zimanyi}}, \bibinfo {author} {\bibfnamefont {B.~A.}\ \bibnamefont
			{Allgood}},\ and\ \bibinfo {author} {\bibfnamefont {G.}~\bibnamefont
			{Blatter}},\ }\bibfield  {title} {\bibinfo {title} {Density of states and
			critical behavior of the coulomb glass},\ }\href@noop {} {\bibfield
		{journal} {\bibinfo  {journal} {Phys. Rev. Lett.}\ }\textbf {\bibinfo
			{volume} {102}},\ \bibinfo {pages} {067205} (\bibinfo {year}
		{2009})}\BibitemShut {NoStop}%
	\bibitem [{\citenamefont {M{\"o}bius}\ and\ \citenamefont
		{Richter}(2010)}]{mobius2010comment}%
	\BibitemOpen
	\bibfield  {author} {\bibinfo {author} {\bibfnamefont {A.}~\bibnamefont
			{M{\"o}bius}}\ and\ \bibinfo {author} {\bibfnamefont {M.}~\bibnamefont
			{Richter}},\ }\bibfield  {title} {\bibinfo {title} {Comment on “density of
			states and critical behavior of the coulomb glass”},\ }\href@noop {}
	{\bibfield  {journal} {\bibinfo  {journal} {Phys. Rev. Lett.}\ }\textbf
		{\bibinfo {volume} {105}},\ \bibinfo {pages} {039701} (\bibinfo {year}
		{2010})}\BibitemShut {NoStop}%
	\bibitem [{\citenamefont {Sarvestani}\ \emph {et~al.}(1995)\citenamefont
		{Sarvestani}, \citenamefont {Schreiber},\ and\ \citenamefont
		{Vojta}}]{sarvestani1995coulomb}%
	\BibitemOpen
	\bibfield  {author} {\bibinfo {author} {\bibfnamefont {M.}~\bibnamefont
			{Sarvestani}}, \bibinfo {author} {\bibfnamefont {M.}~\bibnamefont
			{Schreiber}},\ and\ \bibinfo {author} {\bibfnamefont {T.}~\bibnamefont
			{Vojta}},\ }\bibfield  {title} {\bibinfo {title} {Coulomb gap at finite
			temperatures},\ }\href@noop {} {\bibfield  {journal} {\bibinfo  {journal}
			{Phys. Rev. B}\ }\textbf {\bibinfo {volume} {52}},\ \bibinfo {pages} {R3820}
		(\bibinfo {year} {1995})}\BibitemShut {NoStop}%
	\bibitem [{\citenamefont {Grunewald}\ \emph {et~al.}(1982)\citenamefont
		{Grunewald}, \citenamefont {Pohlmann}, \citenamefont {Schweitzer},\ and\
		\citenamefont {Wurtz}}]{grunewald1982mean}%
	\BibitemOpen
	\bibfield  {author} {\bibinfo {author} {\bibfnamefont {M.}~\bibnamefont
			{Grunewald}}, \bibinfo {author} {\bibfnamefont {B.}~\bibnamefont {Pohlmann}},
		\bibinfo {author} {\bibfnamefont {L.}~\bibnamefont {Schweitzer}},\ and\
		\bibinfo {author} {\bibfnamefont {D.}~\bibnamefont {Wurtz}},\ }\bibfield
	{title} {\bibinfo {title} {Mean field approach to the electron glass},\
	}\href@noop {} {\bibfield  {journal} {\bibinfo  {journal} {Journal of Physics
				C: Solid State Physics}\ }\textbf {\bibinfo {volume} {15}},\ \bibinfo {pages}
		{L1153} (\bibinfo {year} {1982})}\BibitemShut {NoStop}%
	\bibitem [{\citenamefont {Pollak}\ and\ \citenamefont
		{Ortuno}(1982)}]{pollak1982coulomb}%
	\BibitemOpen
	\bibfield  {author} {\bibinfo {author} {\bibfnamefont {M.}~\bibnamefont
			{Pollak}}\ and\ \bibinfo {author} {\bibfnamefont {M.}~\bibnamefont
			{Ortuno}},\ }\bibfield  {title} {\bibinfo {title} {Coulomb interactions in
			anderson localized disordered systems},\ }\href@noop {} {\bibfield  {journal}
		{\bibinfo  {journal} {Solar Energy Materials}\ }\textbf {\bibinfo {volume}
			{8}},\ \bibinfo {pages} {81} (\bibinfo {year} {1982})}\BibitemShut {NoStop}%
	\bibitem [{\citenamefont {Pollak}(1984)}]{pollak1984non}%
	\BibitemOpen
	\bibfield  {author} {\bibinfo {author} {\bibfnamefont {M.}~\bibnamefont
			{Pollak}},\ }\bibfield  {title} {\bibinfo {title} {Non-ergodic behaviour of
			anderson insulators with and without coulomb interactions},\ }\href@noop {}
	{\bibfield  {journal} {\bibinfo  {journal} {Philosophical Magazine B}\
		}\textbf {\bibinfo {volume} {50}},\ \bibinfo {pages} {265} (\bibinfo {year}
		{1984})}\BibitemShut {NoStop}%
	\bibitem [{\citenamefont {Grannan}\ and\ \citenamefont
		{Clare}(1994)}]{grannan1994grannan}%
	\BibitemOpen
	\bibfield  {author} {\bibinfo {author} {\bibfnamefont {E.~R.}\ \bibnamefont
			{Grannan}}\ and\ \bibinfo {author} {\bibfnamefont {C.~Y.}\ \bibnamefont
			{Clare}},\ }\bibfield  {title} {\bibinfo {title} {Grannan and yu reply},\
	}\href@noop {} {\bibfield  {journal} {\bibinfo  {journal} {Phys. Rev. Lett.}\
		}\textbf {\bibinfo {volume} {73}},\ \bibinfo {pages} {2934} (\bibinfo {year}
		{1994})}\BibitemShut {NoStop}%
	\bibitem [{\citenamefont {Mueller}\ and\ \citenamefont
		{Ioffe}(2004)}]{mueller2004glass}%
	\BibitemOpen
	\bibfield  {author} {\bibinfo {author} {\bibfnamefont {M.}~\bibnamefont
			{Mueller}}\ and\ \bibinfo {author} {\bibfnamefont {L.}~\bibnamefont
			{Ioffe}},\ }\bibfield  {title} {\bibinfo {title} {Glass transition and the
			coulomb gap in electron glasses},\ }\href@noop {} {\bibfield  {journal}
		{\bibinfo  {journal} {Phys. Rev. Lett.}\ }\textbf {\bibinfo {volume} {93}},\
		\bibinfo {pages} {256403} (\bibinfo {year} {2004})}\BibitemShut {NoStop}%
	\bibitem [{\citenamefont {Pastor}\ and\ \citenamefont
		{Dobrosavljevi{\'c}}(1999)}]{pastor1999melting}%
	\BibitemOpen
	\bibfield  {author} {\bibinfo {author} {\bibfnamefont {A.}~\bibnamefont
			{Pastor}}\ and\ \bibinfo {author} {\bibfnamefont {V.}~\bibnamefont
			{Dobrosavljevi{\'c}}},\ }\bibfield  {title} {\bibinfo {title} {Melting of the
			electron glass},\ }\href@noop {} {\bibfield  {journal} {\bibinfo  {journal}
			{Phys. Rev. Lett.}\ }\textbf {\bibinfo {volume} {83}},\ \bibinfo {pages}
		{4642} (\bibinfo {year} {1999})}\BibitemShut {NoStop}%
	\bibitem [{\citenamefont {Pankov}\ and\ \citenamefont
		{Dobrosavljevi{\'c}}(2005)}]{pankov2005nonlinear}%
	\BibitemOpen
	\bibfield  {author} {\bibinfo {author} {\bibfnamefont {S.}~\bibnamefont
			{Pankov}}\ and\ \bibinfo {author} {\bibfnamefont {V.}~\bibnamefont
			{Dobrosavljevi{\'c}}},\ }\bibfield  {title} {\bibinfo {title} {Nonlinear
			screening theory of the coulomb glass},\ }\href@noop {} {\bibfield  {journal}
		{\bibinfo  {journal} {Phys. Rev. Lett.}\ }\textbf {\bibinfo {volume} {94}},\
		\bibinfo {pages} {046402} (\bibinfo {year} {2005})}\BibitemShut {NoStop}%
	\bibitem [{\citenamefont {Mueller}\ and\ \citenamefont
		{Pankov}(2007)}]{mueller2007mean}%
	\BibitemOpen
	\bibfield  {author} {\bibinfo {author} {\bibfnamefont {M.}~\bibnamefont
			{Mueller}}\ and\ \bibinfo {author} {\bibfnamefont {S.}~\bibnamefont
			{Pankov}},\ }\bibfield  {title} {\bibinfo {title} {Mean-field theory for the
			three-dimensional coulomb glass},\ }\href@noop {} {\bibfield  {journal}
		{\bibinfo  {journal} {Phys. Rev. B}\ }\textbf {\bibinfo {volume} {75}},\
		\bibinfo {pages} {144201} (\bibinfo {year} {2007})}\BibitemShut {NoStop}%
	\bibitem [{\citenamefont {Bray}\ and\ \citenamefont
		{Moore}(1982)}]{bray1982spin}%
	\BibitemOpen
	\bibfield  {author} {\bibinfo {author} {\bibfnamefont {A.}~\bibnamefont
			{Bray}}\ and\ \bibinfo {author} {\bibfnamefont {M.}~\bibnamefont {Moore}},\
	}\bibfield  {title} {\bibinfo {title} {Spin glasses: the hole story},\
	}\href@noop {} {\bibfield  {journal} {\bibinfo  {journal} {Journal of Physics
				C: Solid State Physics}\ }\textbf {\bibinfo {volume} {15}},\ \bibinfo {pages}
		{2417} (\bibinfo {year} {1982})}\BibitemShut {NoStop}%
	\bibitem [{\citenamefont {Amir}\ \emph {et~al.}(2009)\citenamefont {Amir},
		\citenamefont {Oreg},\ and\ \citenamefont {Imry}}]{amir2009slow}%
	\BibitemOpen
	\bibfield  {author} {\bibinfo {author} {\bibfnamefont {A.}~\bibnamefont
			{Amir}}, \bibinfo {author} {\bibfnamefont {Y.}~\bibnamefont {Oreg}},\ and\
		\bibinfo {author} {\bibfnamefont {Y.}~\bibnamefont {Imry}},\ }\bibfield
	{title} {\bibinfo {title} {Slow relaxations and aging in the electron
			glass},\ }\href@noop {} {\bibfield  {journal} {\bibinfo  {journal} {Phys.
				Rev. Lett.}\ }\textbf {\bibinfo {volume} {103}},\ \bibinfo {pages} {126403}
		(\bibinfo {year} {2009})}\BibitemShut {NoStop}%
	\bibitem [{\citenamefont {Amir}\ \emph {et~al.}(2008)\citenamefont {Amir},
		\citenamefont {Oreg},\ and\ \citenamefont {Imry}}]{amir2008mean}%
	\BibitemOpen
	\bibfield  {author} {\bibinfo {author} {\bibfnamefont {A.}~\bibnamefont
			{Amir}}, \bibinfo {author} {\bibfnamefont {Y.}~\bibnamefont {Oreg}},\ and\
		\bibinfo {author} {\bibfnamefont {Y.}~\bibnamefont {Imry}},\ }\bibfield
	{title} {\bibinfo {title} {Mean-field model for electron-glass dynamics},\
	}\href@noop {} {\bibfield  {journal} {\bibinfo  {journal} {Phys. Rev. B}\
		}\textbf {\bibinfo {volume} {77}},\ \bibinfo {pages} {165207} (\bibinfo
		{year} {2008})}\BibitemShut {NoStop}%
	\bibitem [{\citenamefont {Burin}\ \emph {et~al.}(2006)\citenamefont {Burin},
		\citenamefont {Shklovskii}, \citenamefont {Kozub}, \citenamefont {Galperin},\
		and\ \citenamefont {Vinokur}}]{burin2006many}%
	\BibitemOpen
	\bibfield  {author} {\bibinfo {author} {\bibfnamefont {A.}~\bibnamefont
			{Burin}}, \bibinfo {author} {\bibfnamefont {B.}~\bibnamefont {Shklovskii}},
		\bibinfo {author} {\bibfnamefont {V.}~\bibnamefont {Kozub}}, \bibinfo
		{author} {\bibfnamefont {Y.}~\bibnamefont {Galperin}},\ and\ \bibinfo
		{author} {\bibfnamefont {V.}~\bibnamefont {Vinokur}},\ }\bibfield  {title}
	{\bibinfo {title} {Many electron theory of 1/ f noise in hopping
			conductivity},\ }\href@noop {} {\bibfield  {journal} {\bibinfo  {journal}
			{Phys. Rev. B}\ }\textbf {\bibinfo {volume} {74}},\ \bibinfo {pages} {075205}
		(\bibinfo {year} {2006})}\BibitemShut {NoStop}%
	\bibitem [{\citenamefont {Bhandari}\ \emph {et~al.}(2021)\citenamefont
		{Bhandari}, \citenamefont {Malik}, \citenamefont {Kumar},\ and\ \citenamefont
		{Schechter}}]{bhandari2021relaxation}%
	\BibitemOpen
	\bibfield  {author} {\bibinfo {author} {\bibfnamefont {P.}~\bibnamefont
			{Bhandari}}, \bibinfo {author} {\bibfnamefont {V.}~\bibnamefont {Malik}},
		\bibinfo {author} {\bibfnamefont {D.}~\bibnamefont {Kumar}},\ and\ \bibinfo
		{author} {\bibfnamefont {M.}~\bibnamefont {Schechter}},\ }\bibfield  {title}
	{\bibinfo {title} {Relaxation dynamics of the three-dimensional coulomb glass
			model},\ }\href@noop {} {\bibfield  {journal} {\bibinfo  {journal} {Phys.
				Rev. E}\ }\textbf {\bibinfo {volume} {103}},\ \bibinfo {pages} {032150}
		(\bibinfo {year} {2021})}\BibitemShut {NoStop}%
	\bibitem [{\citenamefont {Kolton}\ \emph {et~al.}(2005)\citenamefont {Kolton},
		\citenamefont {Grempel},\ and\ \citenamefont
		{Dom{\'\i}nguez}}]{kolton2005heterogeneous}%
	\BibitemOpen
	\bibfield  {author} {\bibinfo {author} {\bibfnamefont {A.~B.}\ \bibnamefont
			{Kolton}}, \bibinfo {author} {\bibfnamefont {D.}~\bibnamefont {Grempel}},\
		and\ \bibinfo {author} {\bibfnamefont {D.}~\bibnamefont {Dom{\'\i}nguez}},\
	}\bibfield  {title} {\bibinfo {title} {Heterogeneous dynamics of the
			three-dimensional coulomb glass out of equilibrium},\ }\href@noop {}
	{\bibfield  {journal} {\bibinfo  {journal} {Phys. Rev. B}\ }\textbf {\bibinfo
			{volume} {71}},\ \bibinfo {pages} {024206} (\bibinfo {year}
		{2005})}\BibitemShut {NoStop}%
	\bibitem [{\citenamefont {Kirkengen}\ and\ \citenamefont
		{Bergli}(2009)}]{kirkengen2009slow}%
	\BibitemOpen
	\bibfield  {author} {\bibinfo {author} {\bibfnamefont {M.}~\bibnamefont
			{Kirkengen}}\ and\ \bibinfo {author} {\bibfnamefont {J.}~\bibnamefont
			{Bergli}},\ }\bibfield  {title} {\bibinfo {title} {Slow relaxation and
			equilibrium dynamics in a two-dimensional coulomb glass: Demonstration of
			stretched exponential energy correlations},\ }\href@noop {} {\bibfield
		{journal} {\bibinfo  {journal} {Phys. Rev. B}\ }\textbf {\bibinfo {volume}
			{79}},\ \bibinfo {pages} {075205} (\bibinfo {year} {2009})}\BibitemShut
	{NoStop}%
	\bibitem [{\citenamefont {Vaknin}\ \emph
		{et~al.}(2000{\natexlab{a}})\citenamefont {Vaknin}, \citenamefont
		{Ovadyahu},\ and\ \citenamefont {Pollak}}]{vaknin2000aging}%
	\BibitemOpen
	\bibfield  {author} {\bibinfo {author} {\bibfnamefont {A.}~\bibnamefont
			{Vaknin}}, \bibinfo {author} {\bibfnamefont {Z.}~\bibnamefont {Ovadyahu}},\
		and\ \bibinfo {author} {\bibfnamefont {M.}~\bibnamefont {Pollak}},\
	}\bibfield  {title} {\bibinfo {title} {Aging effects in an anderson
			insulator},\ }\href@noop {} {\bibfield  {journal} {\bibinfo  {journal} {Phys.
				Rev. Lett.}\ }\textbf {\bibinfo {volume} {84}},\ \bibinfo {pages} {3402}
		(\bibinfo {year} {2000}{\natexlab{a}})}\BibitemShut {NoStop}%
	\bibitem [{\citenamefont {Vaknin}\ \emph
		{et~al.}(2000{\natexlab{b}})\citenamefont {Vaknin}, \citenamefont
		{Ovadyahu},\ and\ \citenamefont {Pollak}}]{vaknin2000heuristic}%
	\BibitemOpen
	\bibfield  {author} {\bibinfo {author} {\bibfnamefont {A.}~\bibnamefont
			{Vaknin}}, \bibinfo {author} {\bibfnamefont {Z.}~\bibnamefont {Ovadyahu}},\
		and\ \bibinfo {author} {\bibfnamefont {M.}~\bibnamefont {Pollak}},\
	}\bibfield  {title} {\bibinfo {title} {Heuristic model for slow relaxation of
			excess conductance in electron glasses},\ }\href@noop {} {\bibfield
		{journal} {\bibinfo  {journal} {Phys. Rev. B}\ }\textbf {\bibinfo {volume}
			{61}},\ \bibinfo {pages} {6692} (\bibinfo {year}
		{2000}{\natexlab{b}})}\BibitemShut {NoStop}%
	\bibitem [{\citenamefont {Vaknin}\ \emph {et~al.}(2002)\citenamefont {Vaknin},
		\citenamefont {Ovadyahu},\ and\ \citenamefont
		{Pollak}}]{vaknin2002nonequilibrium}%
	\BibitemOpen
	\bibfield  {author} {\bibinfo {author} {\bibfnamefont {A.}~\bibnamefont
			{Vaknin}}, \bibinfo {author} {\bibfnamefont {Z.}~\bibnamefont {Ovadyahu}},\
		and\ \bibinfo {author} {\bibfnamefont {M.}~\bibnamefont {Pollak}},\
	}\bibfield  {title} {\bibinfo {title} {Nonequilibrium field effect and memory
			in the electron glass},\ }\href@noop {} {\bibfield  {journal} {\bibinfo
			{journal} {Phys. Rev. B}\ }\textbf {\bibinfo {volume} {65}},\ \bibinfo
		{pages} {134208} (\bibinfo {year} {2002})}\BibitemShut {NoStop}%
	\bibitem [{\citenamefont {Orlyanchik}\ and\ \citenamefont
		{Ovadyahu}(2004)}]{orlyanchik2004stress}%
	\BibitemOpen
	\bibfield  {author} {\bibinfo {author} {\bibfnamefont {V.}~\bibnamefont
			{Orlyanchik}}\ and\ \bibinfo {author} {\bibfnamefont {Z.}~\bibnamefont
			{Ovadyahu}},\ }\bibfield  {title} {\bibinfo {title} {Stress aging in the
			electron glass},\ }\href@noop {} {\bibfield  {journal} {\bibinfo  {journal}
			{Phys. Rev. Lett.}\ }\textbf {\bibinfo {volume} {92}},\ \bibinfo {pages}
		{066801} (\bibinfo {year} {2004})}\BibitemShut {NoStop}%
	\bibitem [{\citenamefont {Kozub}\ \emph {et~al.}(2008)\citenamefont {Kozub},
		\citenamefont {Galperin}, \citenamefont {Vinokur},\ and\ \citenamefont
		{Burin}}]{kozub2008memory}%
	\BibitemOpen
	\bibfield  {author} {\bibinfo {author} {\bibfnamefont {V.}~\bibnamefont
			{Kozub}}, \bibinfo {author} {\bibfnamefont {Y.}~\bibnamefont {Galperin}},
		\bibinfo {author} {\bibfnamefont {V.}~\bibnamefont {Vinokur}},\ and\ \bibinfo
		{author} {\bibfnamefont {A.}~\bibnamefont {Burin}},\ }\bibfield  {title}
	{\bibinfo {title} {Memory effects in transport through a hopping insulator:
			Understanding two-dip experiments},\ }\href@noop {} {\bibfield  {journal}
		{\bibinfo  {journal} {Phys. Rev. B}\ }\textbf {\bibinfo {volume} {78}},\
		\bibinfo {pages} {132201} (\bibinfo {year} {2008})}\BibitemShut {NoStop}%
	\bibitem [{\citenamefont {Ben-Chorin}\ \emph {et~al.}(1993)\citenamefont
		{Ben-Chorin}, \citenamefont {Ovadyahu},\ and\ \citenamefont
		{Pollak}}]{ben1993nonequilibrium}%
	\BibitemOpen
	\bibfield  {author} {\bibinfo {author} {\bibfnamefont {M.}~\bibnamefont
			{Ben-Chorin}}, \bibinfo {author} {\bibfnamefont {Z.}~\bibnamefont
			{Ovadyahu}},\ and\ \bibinfo {author} {\bibfnamefont {M.}~\bibnamefont
			{Pollak}},\ }\bibfield  {title} {\bibinfo {title} {Nonequilibrium transport
			and slow relaxation in hopping conductivity},\ }\href@noop {} {\bibfield
		{journal} {\bibinfo  {journal} {Phys. Rev. B}\ }\textbf {\bibinfo {volume}
			{48}},\ \bibinfo {pages} {15025} (\bibinfo {year} {1993})}\BibitemShut
	{NoStop}%
	\bibitem [{\citenamefont {Martinez-Arizala}\ \emph {et~al.}(1997)\citenamefont
		{Martinez-Arizala}, \citenamefont {Grupp}, \citenamefont {Christiansen},
		\citenamefont {Mack}, \citenamefont {Markovic}, \citenamefont {Seguchi},\
		and\ \citenamefont {Goldman}}]{martinez1997anomalous}%
	\BibitemOpen
	\bibfield  {author} {\bibinfo {author} {\bibfnamefont {G.}~\bibnamefont
			{Martinez-Arizala}}, \bibinfo {author} {\bibfnamefont {D.}~\bibnamefont
			{Grupp}}, \bibinfo {author} {\bibfnamefont {C.}~\bibnamefont {Christiansen}},
		\bibinfo {author} {\bibfnamefont {A.}~\bibnamefont {Mack}}, \bibinfo {author}
		{\bibfnamefont {N.}~\bibnamefont {Markovic}}, \bibinfo {author}
		{\bibfnamefont {Y.}~\bibnamefont {Seguchi}},\ and\ \bibinfo {author}
		{\bibfnamefont {A.~M.}\ \bibnamefont {Goldman}},\ }\bibfield  {title}
	{\bibinfo {title} {Anomalous field effect in ultrathin films of metals near
			the superconductor-insulator transition},\ }\href@noop {} {\bibfield
		{journal} {\bibinfo  {journal} {Phys. Rev. Lett.}\ }\textbf {\bibinfo
			{volume} {78}},\ \bibinfo {pages} {1130} (\bibinfo {year}
		{1997})}\BibitemShut {NoStop}%
	\bibitem [{\citenamefont {Martinez-Arizala}\ \emph {et~al.}(1998)\citenamefont
		{Martinez-Arizala}, \citenamefont {Christiansen}, \citenamefont {Grupp},
		\citenamefont {Markovi{\'c}}, \citenamefont {Mack},\ and\ \citenamefont
		{Goldman}}]{martinez1998coulomb}%
	\BibitemOpen
	\bibfield  {author} {\bibinfo {author} {\bibfnamefont {G.}~\bibnamefont
			{Martinez-Arizala}}, \bibinfo {author} {\bibfnamefont {C.}~\bibnamefont
			{Christiansen}}, \bibinfo {author} {\bibfnamefont {D.}~\bibnamefont {Grupp}},
		\bibinfo {author} {\bibfnamefont {N.}~\bibnamefont {Markovi{\'c}}}, \bibinfo
		{author} {\bibfnamefont {A.}~\bibnamefont {Mack}},\ and\ \bibinfo {author}
		{\bibfnamefont {A.}~\bibnamefont {Goldman}},\ }\bibfield  {title} {\bibinfo
		{title} {Coulomb-glass-like behavior of ultrathin films of metals},\
	}\href@noop {} {\bibfield  {journal} {\bibinfo  {journal} {Phys. Rev. B}\
		}\textbf {\bibinfo {volume} {57}},\ \bibinfo {pages} {R670} (\bibinfo {year}
		{1998})}\BibitemShut {NoStop}%
	\bibitem [{\citenamefont {Clare}(1999)}]{clare1999time}%
	\BibitemOpen
	\bibfield  {author} {\bibinfo {author} {\bibfnamefont {C.~Y.}\ \bibnamefont
			{Clare}},\ }\bibfield  {title} {\bibinfo {title} {Time-dependent development
			of the coulomb gap},\ }\href@noop {} {\bibfield  {journal} {\bibinfo
			{journal} {Phys. Rev. Lett.}\ }\textbf {\bibinfo {volume} {82}},\ \bibinfo
		{pages} {4074} (\bibinfo {year} {1999})}\BibitemShut {NoStop}%
	\bibitem [{\citenamefont {Malik}\ and\ \citenamefont
		{Kumar}(2004)}]{malik2004formation}%
	\BibitemOpen
	\bibfield  {author} {\bibinfo {author} {\bibfnamefont {V.}~\bibnamefont
			{Malik}}\ and\ \bibinfo {author} {\bibfnamefont {D.}~\bibnamefont {Kumar}},\
	}\bibfield  {title} {\bibinfo {title} {Formation of the coulomb gap in a
			coulomb glass},\ }\href@noop {} {\bibfield  {journal} {\bibinfo  {journal}
			{Phys. Rev. B}\ }\textbf {\bibinfo {volume} {69}},\ \bibinfo {pages} {153103}
		(\bibinfo {year} {2004})}\BibitemShut {NoStop}%
	\bibitem [{\citenamefont {Lebanon}\ and\ \citenamefont
		{M{\"u}ller}(2005)}]{lebanon2005memory}%
	\BibitemOpen
	\bibfield  {author} {\bibinfo {author} {\bibfnamefont {E.}~\bibnamefont
			{Lebanon}}\ and\ \bibinfo {author} {\bibfnamefont {M.}~\bibnamefont
			{M{\"u}ller}},\ }\bibfield  {title} {\bibinfo {title} {Memory effect in
			electron glasses: Theoretical analysis via a percolation approach},\
	}\href@noop {} {\bibfield  {journal} {\bibinfo  {journal} {Phys. Rev. B}\
		}\textbf {\bibinfo {volume} {72}},\ \bibinfo {pages} {174202} (\bibinfo
		{year} {2005})}\BibitemShut {NoStop}%
	\bibitem [{\citenamefont {Ovadyahu}(2019)}]{ovadyahu2019screening}%
	\BibitemOpen
	\bibfield  {author} {\bibinfo {author} {\bibfnamefont {Z.}~\bibnamefont
			{Ovadyahu}},\ }\bibfield  {title} {\bibinfo {title} {Screening the coulomb
			interaction and thermalization of anderson insulators},\ }\href@noop {}
	{\bibfield  {journal} {\bibinfo  {journal} {Phys. Rev. B}\ }\textbf {\bibinfo
			{volume} {95}},\ \bibinfo {pages} {184201} (\bibinfo {year}
		{2019})}\BibitemShut {NoStop}%
	\bibitem [{\citenamefont {Ovadyahu}(2017)}]{ovadyahu2017disorder}%
	\BibitemOpen
	\bibfield  {author} {\bibinfo {author} {\bibfnamefont {Z.}~\bibnamefont
			{Ovadyahu}},\ }\bibfield  {title} {\bibinfo {title} {Slow dynamics of
			electron glasses: The role of disorder},\ }\href@noop {} {\bibfield
		{journal} {\bibinfo  {journal} {Phys. Rev. B}\ }\textbf {\bibinfo {volume}
			{95}},\ \bibinfo {pages} {134203} (\bibinfo {year} {2017})}\BibitemShut
	{NoStop}%
	\bibitem [{\citenamefont {Puri}\ and\ \citenamefont
		{Wadhawan}(2009)}]{Puri2009}%
	\BibitemOpen
	\bibfield  {author} {\bibinfo {author} {\bibfnamefont {S.}~\bibnamefont
			{Puri}}\ and\ \bibinfo {author} {\bibfnamefont {V.}~\bibnamefont
			{Wadhawan}},\ }\href@noop {} {\emph {\bibinfo {title} {Kinetics of phase
				transitions}}}\ (\bibinfo  {publisher} {CRC press},\ \bibinfo {year}
	{2009})\BibitemShut {NoStop}%
	\bibitem [{\citenamefont {Ovadyahu}(2018)}]{ovadyahu2018transition}%
	\BibitemOpen
	\bibfield  {author} {\bibinfo {author} {\bibfnamefont {Z.}~\bibnamefont
			{Ovadyahu}},\ }\bibfield  {title} {\bibinfo {title} {Transition to
			exponential relaxation in weakly disordered electron glasses},\ }\href@noop
	{} {\bibfield  {journal} {\bibinfo  {journal} {Phys. Rev. B}\ }\textbf
		{\bibinfo {volume} {97}},\ \bibinfo {pages} {214201} (\bibinfo {year}
		{2018})}\BibitemShut {NoStop}%
	\bibitem [{\citenamefont {Givan}\ and\ \citenamefont {Ovadyahu}(2012)}]{givan}%
	\BibitemOpen
	\bibfield  {author} {\bibinfo {author} {\bibfnamefont {U.}~\bibnamefont
			{Givan}}\ and\ \bibinfo {author} {\bibfnamefont {Z.}~\bibnamefont
			{Ovadyahu}},\ }\bibfield  {title} {\bibinfo {title} {Compositional disorder
			and transport peculiarities in the amorphous indium oxides},\ }\href@noop {}
	{\bibfield  {journal} {\bibinfo  {journal} {Phys. Rev. B}\ }\textbf {\bibinfo
			{volume} {86}},\ \bibinfo {pages} {165101} (\bibinfo {year}
		{2012})}\BibitemShut {NoStop}%
	\bibitem [{\citenamefont {Ovadyahu}\ and\ \citenamefont
		{Pollak}(1997)}]{ovadyahu1997disorder}%
	\BibitemOpen
	\bibfield  {author} {\bibinfo {author} {\bibfnamefont {Z.}~\bibnamefont
			{Ovadyahu}}\ and\ \bibinfo {author} {\bibfnamefont {M.}~\bibnamefont
			{Pollak}},\ }\bibfield  {title} {\bibinfo {title} {Disorder and magnetic
			field dependence of slow electronic relaxation},\ }\href@noop {} {\bibfield
		{journal} {\bibinfo  {journal} {Phys. Rev. Lett.}\ }\textbf {\bibinfo
			{volume} {79}},\ \bibinfo {pages} {459} (\bibinfo {year} {1997})}\BibitemShut
	{NoStop}%
\end{thebibliography}
%

\end{document}